\documentclass[nocopyright]{cimento}
\usepackage{url}
\usepackage{gensymb}
\usepackage{physics}
\usepackage{upgreek}

\title{Dark matter and fundamental physics with the Cherenkov Telescope Array Observatory}
\shorttitle{Dark matter and fundamental physics with CTAO}

\author{F.~Schiavone\thanks{francesco.schiavone@ba.infn.it} for the CTAO Consortium}
\shortauthor{F.~Schiavone}

\instlist{
\inst{} Dipartimento Interateneo di Fisica ``Michelangelo Merlin'', Università degli Studi di Bari and INFN Sezione di Bari
    }

\begin{document}

\maketitle

\begin{abstract}
The Cherenkov Telescope Array Observatory (CTAO) will be the next-generation major ground-based gamma-ray observatory. It will be made up of two large arrays of imaging atmospheric Cherenkov telescopes (IACTs), with one site in the Northern hemisphere (La Palma, Canary Islands) and one in the Southern (Paranal, Chile). CTAO aims to offer great improvement in energetic and angular resolution with respect to current IACT systems, spanning a photon energy range from 20 GeV to 300 TeV, as well as a significantly larger effective area and full-sky coverage. Besides a percentage of observational time available for external proposals, making it the first open gamma-ray observatory, the core observational program of CTAO is organized in several Key Science Projects. A significant amount of time will be devoted to dark matter (e.g. WIMPs, axion-like particles) and fundamental physics studies from a variety of targets, including the Milky Way's Galactic centre, dwarf spheroidal galaxies, the Large Magellanic Cloud, and extragalactic objects such as blazars. In this contribution, an overview of CTAO’s main features is provided, with a focus on its capabilities to investigate these yet unanswered questions of modern physics.
\end{abstract}

\section{Gamma-ray astrophysics and Cherenkov telescopes}
The field of gamma-ray astrophysics was born around the 1950s, when astronomers and particle physicists began developing techniques to detect gamma rays of cosmic origin \cite{hillas2013}. The aim of these searches was to characterize high-energy processes happening in astrophysical objects, such as the acceleration of cosmic rays that were abundantly measured at ground level or by balloon experiments. The advantage of gamma rays in this regard would be the fact that, unlike charged particles, they could not be deflected by magnetic fields, such as those present in the Milky Way or between galaxies, and would thus directly point to the sources of emission. 

Astrophysical gamma rays are quickly absorbed by the Earth's atmosphere, as their interactions with the atmospheric nuclei causes pair production of relativistic electrons and positrons. These can produce further gamma-ray photons of lower energies via bremsstrahlung, thus giving rise to atmospheric showers. For this reason, a direct measurement of cosmic gamma rays from the Earth's surface is impossible. A successful approach to overcome this limitation is represented by satellite-based telescopes, such as the recently decommissioned AGILE\footnote{\url{http://agile.rm.iasf.cnr.it/}} and the widely successful \textit{Fermi} Gamma-Ray Space Telescope\footnote{\url{https://fermi.gsfc.nasa.gov/}}, equipped with the Large Area Telescope (LAT) and Gamma-ray Burst Monitor (GBM) instruments. Space-borne detectors, however, have relatively small effective areas (around $\rm 1\,m^2$) which represent a strong statistical limitation in measuring the very low gamma-ray fluxes above some hundreds of GeV.

In order to access higher energies in the electromagnetic spectrum, different ground-based approaches have been developed. One of such approaches aims to infer the cosmic gamma-ray's direction and energy by measuring the Cherenkov light produced by the relativistic charged particles in the electromagnetic showers, effectively using the atmosphere as a ``natural calorimeter''. The light from showers is emitted in a cone, 
resulting in a ``light pool'' of almost constant intensity at ground level with a radius of about $100\,\rm m$ \cite{denaurois2015}. In this way, the effective area for detecting gamma rays becomes of order $\rm10^4\,m^2$, making it possible to measure much smaller fluxes and, thus, higher energies. 

This is the basic working principle of imaging atmospheric Cherenkov telescopes (IACTs). The Cherenkov light is collected by large reflectors and recorded by cameras using photomultiplier tubes (PMTs) or, more recently, silicon photomultipliers (SiPMs). The shapes of the shower images are then used to reconstruct the properties of the original gamma rays, and to discriminate the background due to the much more abundant charged cosmic rays. There are three major IACT systems currently in operation: MAGIC\footnote{\url{https://magic.mpp.mpg.de/}}, at Observatorio del Roque de los Muchachos in La Palma, Canary Islands, made up of two $\rm 17\,m$ reflectors; H.E.S.S.\footnote{\url{https://www.mpi-hd.mpg.de/HESS/}}, on the Khomas highlands in Namibia, featuring four $\rm 12\,m$ reflectors along with a larger  one with $\rm28\,m$ diameter; and VERITAS\footnote{\url{https://veritas.sao.arizona.edu/}}, at the Fred Lawrence Whipple Observatory in Arizona, US, with four $\rm 12\,m$ dishes. In all cases, multiple telescopes are needed to improve background rejection and stereoscopic reconstruction of gamma-ray events.

\section{The Cherenkov Telescope Array Observatory}
The future of IACTs will be represented by the Cherenkov Telescope Array Observatory (CTAO)\footnote{\url{https://www.ctao.org/}}, which aims to improve the energetic and spatial resolution with respect current Cherenkov telescopes, as well as to offer full coverage of the night sky. Additionally, CTAO plans to extend the energy range to energies as low as $\rm20\,GeV$, to overlap with \textit{Fermi} and similar instruments, and as high as $\rm300\,TeV$, to probe the most energetic Galactic sources. 

In order to cover the energy range required, three different types of telescopes have been designed \cite{acharya2013}. Few Large-Sized Telescopes (LSTs), with a 23-metre diameter and $4.3\degree$ field of view (FoV), are able to image the frequent, feeble low-energy showers (20--150$\rm\,GeV$), while many Small-Sized Telescopes (SSTs), with $\rm4.3\,m$ primary reflectors and $9\degree$ FoV, are most sensitive to the rare but very bright showers in the 5--300$\rm\,TeV$ range. Finally, 12-metre Medium-Sized Telescopes (MSTs) with a $8\degree$ FoV are dedicated to the core energies, from $\rm150\,GeV$ to $\rm5\,TeV$. 

Two large telescope arrays are planned in the Northern and Southern hemispheres. The Northern Array, which will be located at Observatorio del Roque de los Muchachos in La Palma, will be made up of 4 LSTs and 9 MSTs and will be mainly dedicated to observe the gamma-ray emission of extragalactic sources up to a few TeV. The Southern Array, in the Atacama desert, Chile, will be optimized to detect the highest-energy photons from the inner regions of the Milky Way, and will consist of 14 MSTs and 37 SSTs.

Working prototypes of each kind of telescope already exist. Most notably LST-1\footnote{\url{https://www.lst1.iac.es/}}, the first LST of the CTAO Northern Array, is currently ending its commissioning phase and already producing scientific results \cite{abe2023-lst-a, abe2023-lst-b, abe2024-lst-a, abe2025-lst-a, abe2025-lst-b}, while the other three Northern LSTs are approaching completion. Advanced prototypes of a dual-mirror design for the SSTs, designed within the ASTRI\footnote{\url{http://www.astri.inaf.it/?lang=en}} project, are currently in operation at the Serra La Nave observatory, Catania, and at Observatorio del Teide in Tenerife. In addition, MST and single-mirror SST prototypes have been installed in Berlin and Prague, respectively. Foreseen upgrades to the Southern Array include the construction of two LSTs and five additional SSTs within the Italian PNRR CTA+ Project\footnote{\url{https://pnrr.inaf.it/progetto-ctaplus/}}. An alternative dual-mirror design for the medium-sized telescopes (Schwarzschild-Couder Telescope, SCT) has been proposed as well, with a working prototype at the Whipple Observatory in Arizona \cite{adams2021}.

Finally, CTAO will be the first ground-based gamma-ray observatory open to proposals from external investigators. However, in the first ten years of its operation, 40\% of its observational time will be devoted to different Key Science Projects (KSPs) defined by the CTAO Consortium \cite{ctascience2018}. Among these, a significant amount of observations will be dedicated to dark matter searches from a variety of targets and fundamental physics studies. Such an interest is also demonstrated by several papers published by the Consortium in the recent years, which we review below.

\section{Indirect dark matter searches with CTAO}
It is well-known that about 32\% of the energy density of the Universe consists of non-relativistic matter \cite{aghanim2020}. Various observational pieces of evidence show that only about 15\% of this fraction is visible in the form of baryonic matter, i.e. astrophysical objects such as stars, galaxies and gas clouds, while the rest must be found in some form of gravitationally-interacting ``cold dark matter''. While the precise particle nature of dark matter is still unknown, it should consist of particles which are either stable or with a decay time comparable to the age of the Universe, neutrally charged with respect to all fundamental interactions except for gravity, and whose presently observed abundance can be explained by some cosmological mechanism. Some of the best-motivated candidates fulfilling these requirements are represented by weakly-interacting massive particles (WIMPs). According to the WIMP paradigm, in the early Universe these particles would be able to self-annihilate into pairs of Standard Model particles, and vice versa. As a consequence of cosmological expansion, this process would gradually go out of equilibrium, leaving a relic WIMP abundance in a ``freeze-out'' process \cite{kolb1990}. The reference value for the velocity-averaged annihilation cross section is given by the thermal estimate $\langle\sigma v\rangle\simeq3\times10^{-26}{\,\rm cm^3\,s^{-1}}$, which reproduces the present dark matter density.

In gamma-ray astrophysics, WIMP searches are usually performed by looking for signatures of annihilation or decay in the spectra of massive targets, where the dark matter density is expected to be highest. In the case of annihilation, the expected contribution to the gamma-ray flux is given by
\begin{equation}
    \dv{\Phi}{E}=\frac{1}{8\pi}\frac{\langle\sigma v\rangle}{m^2_\chi}\dv{N_\gamma}{E}\times\int_{\Delta\Omega}\int_{\rm l.o.s.}\rho^2(\ell)\dd{\ell}\,,
\end{equation}
which is conventionally split in two factors. The particle physics factor involves the WIMP mass $m_\chi$, the velocity-averaged annihilation cross-section $\langle\sigma v\rangle$ and the resulting photon spectrum $\dv*{N_\gamma}{E}$, which depends on the specific particle physics model used. Typically, $\dv*{N_\gamma}{E}$ is computed considering annihilation into a few benchmark channels such as $W^+W^-$, $\tau^+\tau^-$ or $b\bar{b}$ pairs \cite{cirelli2011}. The second factor, usually called astrophysical factor or $J$-factor, contains the squared dark matter density profile $\rho(\ell)$ integrated over the line of sight from Earth and the solid angle $\Delta\Omega$ covered by the detector. Common theoretically motivated density profiles are those by Einasto and by Navarro, Frenk and White (NFW).

Searches of dark matter in the form of WIMPs represent one of the main KSPs of CTAO, involving multiple sources, for which several hundreds of hours are expected to be allocated. A paramount target for this kind of studies is the Galactic Center of the Milky Way, as its large mass and relative proximity to Earth would guarantee a strong dark matter signal. However, such a signal should be disentangled from the equally large background from the many gamma-ray sources present in this region. This, in turn, requires both an accurate modeling of all its components, and a high enough energy and spatial resolution, which is one of CTAO's main improvements with respect to current IACT systems. Based on the KSP specifications, observations of the Galactic Center and surrounding regions have been simulated for a total of 825 hours, demonstrating that CTAO should be able to improve the current upper limits on $\langle\sigma v\rangle$ for TeV-scale WIMPs by about one order of magnitude within its first 10 years \cite{acharyya2021}. Similar studies were performed by simulating observations of the Large Magellanic Cloud for 340 hours \cite{acharyya2023} and for the Perseus Galaxy Cluster for 300 hours \cite{abe2024a}, where the case of WIMP decay was considered as well to place lower-limits on the WIMP lifetime. While these targets would present a weaker dark matter signal, the background contamination would also be significantly more moderate, allowing for an easier detection.

A different approach to dark matter searches is implemented by looking for specific gamma-ray spectral features, e.g. monochromatic lines from direct WIMP annihilation into photon pairs, irrespective of the underlying particle physics model. In this case, simulated observations of the Galactic Center (500 hours) and of six dwarf spheroidal galaxies (100 hours each) with CTAO show that upper limits on $\langle\sigma v\rangle$ for TeV-scale WIMPs can be set even several orders of magnitude below the thermal benchmark \cite{abe2024b}.

\section{Axion-like particles and other fundamental physics}
Gamma-ray observations with CTAO will also be used to investigate questions related to cosmology and fundamental physics. Contrary to dark matter searches, which benefit from the relative proximity of the sources, this kind of studies aims to constrain effects on the propagation of gamma-ray photons over cosmological distances. For this reason, a privileged target is represented by gamma-ray emitting active galactic nuclei (AGNs). These extragalactic sources are located at a distance from the Milky Way measured in terms of their cosmological redshift parameter $z$, with typical values between 0 and 2, which permits to study non-standard photon propagation effects. The CTAO AGN KSP aims to perform in-depth observations of selected AGNs from different classes, as well as a monitoring campaign in search of high-intensity flares.

An important topic of research is represented by axion-like particles (ALPs). A generalization of the axion, originally introduced in the context of quantum chromodynamics as a solution to the strong CP problem \cite{pecceiquinn1977, weinberg1978, wilczek1978}, ALPs are a common prediction of many beyond-Standard-Model theories such as string theory, and are often quoted as an alternative to WIMPs as possible constituents of dark matter. They are typically characterised by two parameters, a mass $m_a$ and an effective two-photon vertex whose strength is measured by the coupling constant $g_{a\gamma}$. The latter property allows for ALP-photon oscillations in the presence of external magnetic fields, which, for astrophysical objects, translates into a modification of the observed gamma-ray flux. This can be expressed as
\begin{equation}
    \dv{\Phi_{\rm obs}}{E}=P_{\gamma\gamma}\dv{\Phi_{\rm int}}{E}\,,
\end{equation}
where $\dv*{\Phi_{\rm obs}}{E}$ and $\dv*{\Phi_{\rm int}}{E}$ are the observed and intrinsic photon fluxes, respectively, while $P_{\gamma\gamma}$ is a photon survival probability sensitive to the photon energy, ALP parameters $m_a$ and $g_{a\gamma}$, source distance and magnetic field intensity and morphology along the line of sight.

Simulated observations of the AGN NGC~1275, located at the center of the Perseus Galaxy Cluster, yield an estimate of the capability of CTAO to constrain the ALP parameter space \cite{abdalla2021}. These results were produced by considering a 300-hour exposure for a quiescent state of the source (overlapping with the Perseus Galaxy Cluster KSP) and a 10-hour exposure for a flaring state. Magnetic fields both inside the galaxy cluster (modeled as a turbulent field with an average intensity of a few $\rm\upmu G$) and the Milky Way were considered. The resulting limits, which are the only ones currently published by the CTAO Consortium, are not very competitive with previous constraints. There are hints, however, that these results could be improved by a significant margin by considering different sources or applying alternative analysis techniques \cite{schiavone2025}.

Simulated AGN observations were used to estimate CTAO's sensitivity to several other effects related to cosmological photon propagation, including measurements of the extragalactic bacgkround light, constraints on the intensity of intergalactic magnetic fields and probes of Lorentz invariance violation. The interested reader is referred to the cited paper \cite{abdalla2021}.

\section{Conclusions}
CTAO will be a revolutionary instrument, improving upon every aspect of performance with respect to current IACTs. For the first time it will offer the possibility to carry out extended surveys for discovering new gamma-ray sources at multi-TeV energies, while observing known targets with unprecedented flux sensitivity and energetic and spatial resolution.

Besides more conventional astrophysics, dark matter searches will play a crucial role in defining the scientific program of the Observatory. Simulated observations of several sources, namely the Milky Way's Galactic Center, the Perseus Galaxy Cluster, the Large Magellanic Cloud and dwarf spheroidal galaxies have shown that CTAO will be able to provide some of the strongest constraints available on TeV-scale dark matter. Similarly, different scenarios in the context of fundamental and exotic physics will be explored by studying effects on photon propagation over cosmological distances, which will be possible thanks to the observation of AGNs. 

\acknowledgments
CTAO gratefully acknowledges financial support from the agencies and organizations listed at \url{https://www.ctao.org/for-scientists/library/acknowledgments/}.

\bibliographystyle{varenna}
\bibliography{biblio.bib}

\end{document}